**TITLE:**

Modelling fertility potential in survivors of childhood cancer: An introduction to modern statistical and computational methods

**AUTHORS AND AFFILIATIONS:**

Lin Yu[1,*], Zhe Lu[1,*], Paul C. Nathan[2], Sogol Mostoufi-Moab[3], Yan Yuan[1]

[1] School of Public Health, University of Alberta, Edmonton, AB, Canada
[2] The Hospital for Sick Children, Toronto, ON, Canada
[3] University of Pennsylvania, Perelman School of Medicine, and The Children's Hospital of Philadelphia, Philadelphia, PA, USA
*These authors contributed equally.

Email addresses of co-authors:
Lin Yu                (lyu6@ualberta.ca)
Zhe Lu                (zhe1@ualberta.ca)
Paul C. Nathan        (Paul.Nathan@sickkids.ca)
Sogol Mostoufi-Moab   (MOAB@email.chop.edu)

Corresponding author:
Yan Yuan              (yyuan@ualberta.ca)

**KEYWORDS:**
Predictive modelling, Logistic regression, Support Vector Machines, Random Forest

**SUMMARY:**
Using an example of predicting female fertility potential post cancer treatment, the study illustrates how statistical and computational methods are conducted. Specifically, we discuss data preparation, algorithm development, and performance assessment in the modelling process. It aims to be a reference for using statistical and computational methods in medical research.

**ABSTRACT:**
Statistical and computational methods are widely used in today's scientific studies. Using a female fertility potential in childhood cancer survivors as an example, we illustrate how these methods can be used to extract insight regarding biological processes from noisy observational data in order to inform decision making. We start by contextualizing the computational methods with the working example: the modelling of acute ovarian failure risk in female childhood cancer survivors to quantify the risk of permanent ovarian failure due to exposure to lifesaving but nonetheless toxic cancer treatments. This is followed by a description of the general framework of classification problems. We provide an overview of the modelling algorithms employed in our example, including one classic model (logistic regression) and two popular modern learning methods (random forest and support vector machines). Using the working example, we show the general steps of data preparation for modelling, variable selection steps for the classic model, and how model performance might be improved utilizing visualization tools. We end with a note on the importance of model evaluation.

**INTRODUCTION:**
Prediction of disease occurrence and progression is important in clinical practice to inform decision-making and patient counselling. Statistical and computational methods are a key part in this process.[1] In this article, we describe the basic framework and steps in developing a prediction algorithm using established computational methods. Specifically, we discuss the steps in the process, including data preparation, algorithm development, predictor selection, and performance assessment. It aims to be a reference for using statistical and computational methods for risk prediction in medical research. We start by contextualizing the computational methods with a working example, the modeling of acute ovarian failure (AOF) risk in childhood cancer survivors to quantify the risk of permanent loss of ovarian function due to exposure to lifesaving but nonetheless toxic cancer treatments.

The decline in the ovarian follicle pool is associated with reduced fertility as a female ages. Menopause normally occurs in the early fifties.[2] However, cancer treatment can accelerate the process of a declining ovarian reserve, resulting in menopause prior to age 40, also known as premature ovarian insufficiency (POI).[3] AOF is a type of POI: it occurs when an individual permanently stops menstruating within five years of their cancer diagnosis or fails to achieve menarche by 18 years of age following cancer treatment.[4]

AOF status is binary. Similar to many other health conditions, such as diabetes, a person is classified as either having the condition or not having the condition. Being able to predict an individual patient's risk of developing these conditions is of particular clinical interest, allowing the provision of proper care for the patient. In data science, these types of variables are termed binary outcomes. The task of modeling a binary outcome is known as classification. It is an important and well-recognized problem widely encountered in many research areas, including medical science, engineering, pattern recognition, etc.[5,6] In the last three decades, many classification algorithms have been developed, thanks to the exponential growth of computing power and the flourishing data science community. In this article, we introduce three classification algorithms, namely logistic regression,[7] support vector machines[8] and random forest[9] to predict the AOF risk, and key criteria for evaluating prediction. We begin with a summary of these three methods and the evaluation metrics.

Logistic regression is a classic classification method and one of the most popular approaches for modeling the relationship between multiple predictors and a binary outcome[1]. It can be used to: 1) estimate the probability of the occurrence of the outcome; 2) explore the risk factors for the outcome, and determine the relative importance of risk factors with respect to the outcome; 3) analyze potential interactions between predictors and control for confounding factors[7]. The general form of a logistic regression model is $ln(p/(1-p)) = \beta_0 + \beta_1 X_1 + \beta_2 X_2 \ldots + \beta_k X_k$, where p denotes the probability of the outcome occurring and p/(1-p) is the odds of the outcome. $ln$ denotes the natural logarithm. $X_1, X_2, \ldots X_k$ are the predictors and $\beta_1, \beta_2, \ldots, \beta_k$ are their coefficients, and $\beta_0$ is the intercept. Logistic regression models the log odds of an outcome using a linear combination of predictors. Logistic regression has been extended to handle multicategory outcomes, i.e., multinomial logistic regression and ordinal logistic regression.[10]

Introduced in 1995, support vector machines (SVM) is a relatively new algorithm[6]. The support vector machine aims to construct a hyperplane that separates the space defined by the projection of the predictors into two parts based on the yes/no label of the outcome from the sample data. A new patient can then be assigned to positive (case) or negative (non-case) based on her predictors' values using this hyperplane. The SVM is efficient as it only depends on patients "located" on or near the margin of the decision boundary.[11] The simplest case is that the patients are separable via a hyperplane defined by a linear combination of the predictors, which is also known as the maximal margin classifier. The maximal margin classifier can be extended to allow some noise, i.e. positive patients with "yes" labels are in the negative part of the space. When a linear boundary cannot separate cases and controls successfully, SVM uses a "kernel", which projects the original predictor space into some higher-dimensional space. This makes it possible to construct a linear hyperplane in the transformed space. However, the hyperplane is nonlinear in the input space.

Proposed in 2001, random forest comprises a large group of decision trees.[9] A decision tree starts from the mixture of the entire sample data of cases and non-cases (known as a root node), recursively selects one predictor among all predictors such that the separation of cases and non-cases is maximized when the node is split into two daughter nodes using the selected predictor. Random forest is a collection of decision trees that are built on bootstrapped samples, i.e., the training set for each tree is a sample with replacement of the original sample data of cases and non-cases. Random forest averages the predicted risks from a large collection of decision trees and is shown to improve the resulting classification precision. One main concern with multiple decision trees is that strong predictors tend to appear at the top of many of these trees. Thus, even when trees are built on different bootstrap samples, these decision trees are often highly similar and correlated.[11] To reduce the correlation among these trees, random forest employs a strategy: randomly select a small number of predictors from all available predictors at each splitting. One best predictor among the small sample of predictors is then selected to split the data as described above.

The goal of AOF risk prediction is to create a clinical tool that gives reliable risk estimates for individual patients. Therefore, the evaluation of the prediction performance of these models is critically important. We discuss three concepts: calibration, discrimination, and predictive ability in the rest of this section. The algorithm should be internally and externally validated before recommended for clinical use.

Calibration measures the magnitude of the predicted probabilities that aligns with the observed outcomes.[12] The calibration curve is often used to evaluate calibration visually. It can be a plot of cumulative observed outcomes against cumulative predicted outcomes. A curve that approximately lies along the y=x line indicates a good agreement between the observed outcome and predicted outcome.

Discrimination measures the ability of the algorithm to discriminate between cases and non-cases. The Area Under the ROC Curve (AUC) is the most commonly used metric for assessing the discrimination of prediction algorithms.[13] The ROC curve is the plot of sensitivity (true positive rate, TPR) against 1-specificity (false positive rate, FPR) at all possible thresholds. The AUC summarizes the performance of the algorithm across all thresholds. The AUC can be interpreted as the probability that the risk score of a randomly selected case is greater than that of a randomly selected non-case. The higher the AUC, the better the discrimination performance of the algorithm.

The prediction ability is measured by the Precision-Recall curve, a plot of precision (also known as positive predictive value) against recall (also known as sensitivity or TPR) for every possible threshold. The area under the precision-recall (PR) curve termed average precision (AP) is the summary metric for prediction ability.[14] Compared to the AUC, the AP gives more weight to the correct classification of cases into the high-risk group. AP is more useful when the proportions of cases and non-cases are highly imbalanced.

The internal validation is typically conducted using cross-validation. The goal of prediction is to apply the predictive model to predict on new data. Therefore, it is important to evaluate the predictive model using external data to ensure external validity.

**PROTOCOL:**

**1. Data preparation**

**1.1 Software environment:**

Install R in a local computer and install necessary R packages. Install "readxl", "haven" and "tidyverse" packages. Import Excel data files and SAS data files into R.

**Note:** Our working example uses the data from the Childhood Cancer Survivor Study (CCSS), which is a longitudinal study of childhood cancer survivors diagnosed between 1970 and 1999[15].

**1.2 Data preprocessing**

1.2.1 Build data documentation in an Excel spreadsheet or Word file. Record important information including data receiving date, contents and number of observations.

1.2.2 Create a data dictionary: For each data set, record variables' names, meaning and types (e.g. "numerical", "factor", or "character"). Create labels for each predictor used in downstream analysis, clearly describe the meaning and type of predictors.

1.2.3 Code predictors: Code predictors in a proper format. For dates, such as birthdate and date of diagnosis, code them as a POSIXct class by using the R builtin function "as.Date". For categorical variables, code them as factor class and specify the levels and labels for factor variables based on the data dictionary.

1.2.4 Logic check: Examine the data for extreme or strange values, e.g. large chemotherapy drug doses which are not consistent with other records, patient age at diagnosis not meeting the inclusion criterion.

**Note:** If raw data is provided by a third party, resolve these data issues with the data provider. The examination of the raw data will not stop at this stage. As we explore data further and understand data better, we should always be careful to check the accuracy and logic of the data.

1.2.5 Exclusion criteria: Exclude participants from datasets if they met any of the exclusion criteria.

**Note:** in this AOF example, the exclusion criteria are: 1) participants were exposed to a pituitary radiation dose higher than 30 Gray; 2) participants had a tumor in the hypothalamus or pituitary region 3) participants' menstrual history information was incomplete, unclear, or provided by a proxy 4) participants experienced a second malignant neoplasm within 5 years of primary cancer diagnosis

1.2.6 Combine data sets: Merge potential predictors from different raw data sets into a single work data set for developing models. For example, in our AOF risk prediction, potential predictors were scattered in 7 raw data sets. Their relationships are presented in an Entity-Relationship Diagram (**Figure 1**).

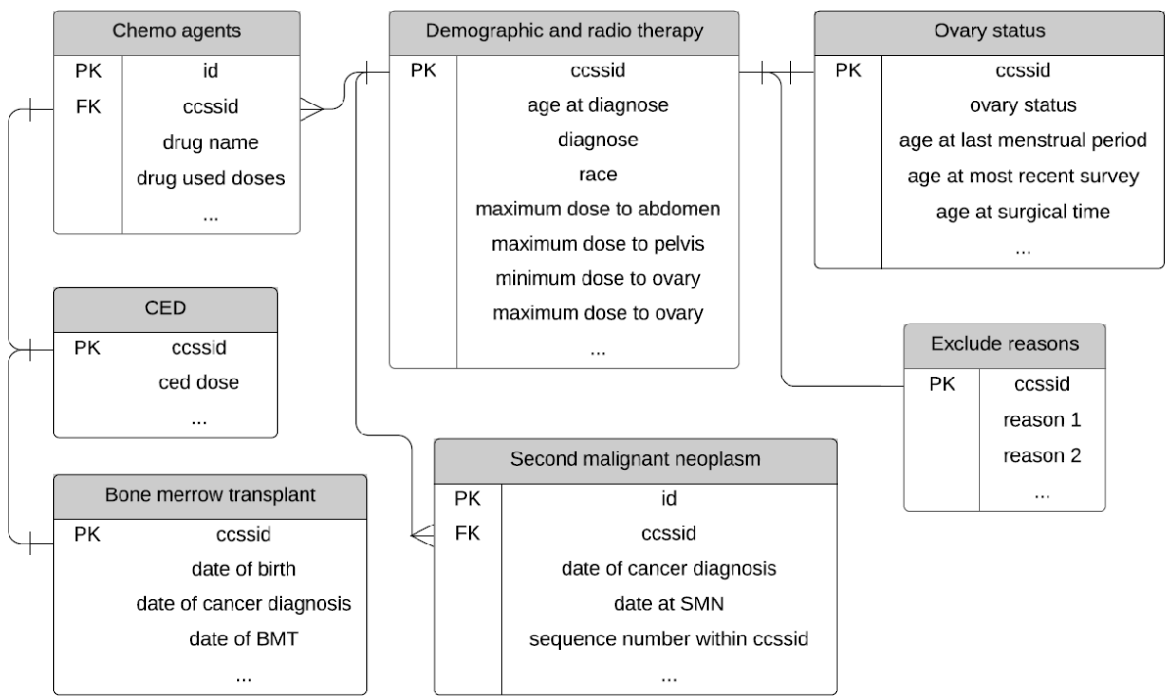

**Figure 1 Entity Relationship Diagrams.** Each box represents a raw data set. PK: "Primary Key" is a field in a table which uniquely identifies each row/record in a data set. FK: "Foreign Key" is a

key used to link two data sets together

**Note:** Before combining them, preprocess the foreign keys to make sure that they have the same name and format. After merging, examine data across the combined data set to make sure information from different raw data sets are consistent.

**1.3 Predictors and outcome list**

Make a list of predictors and outcomes.

In our example, potential predictors include: Age at Cancer Diagnosis, Bone Marrow Transplant (BMT) status, Minimum Ovarian Radiation Dose, Average Ovarian Radiation Dose, Maximum Ovarian Radiation Dose, Cyclophosphamide Equivalence Dose (CED) Value, Cancer Diagnosis Type and Year of Cancer Diagnosis. Outcome is AOF (patients who experienced AOF were coded as 1, otherwise coded as 0).

**2. Algorithm development**

**2.1 Install R packages:**

Install "stats", "e1071", "randomForestSRC" and "APtools" in R.

**Note:** The "glm" function from "stats" package in R was used to fit the logistic regression. The "svm" function from the "e1071" package in R was used to fit the SVM algorithm. The "rfsrc" function from the "randomForestSRC" package in R was used to fit the random forest algorithm. The "APBinary" function from the "APtools" package was used to estimate the AUC values, AP values, and their 95% confidence intervals.

**2.2 Predictor screening:**

Use the criteria of prediction performance to screen predictors and their combinations.

**Note:** In predictive modeling, the goal is to build an algorithm that can generate accurate prediction of the outcome that is unlabeled using predictors. Therefore, instead of focusing on the association between predictors and outcome, we would choose predictors according to the model prediction performance and the applicability of predictors in a real-world setting. In our example, we dropped the Year of Cancer Diagnosis as they will not be applicable in future prediction settings. For identifying interaction, we used two strategies: 1) included clinical plausible interaction effects; 2) using the results from the random forest. For example, the interaction between Age at Cancer Diagnosis and Bone Marrow Transplant (BMT) status was detected by plotting the estimated AOF risk against Age at Cancer Diagnosis stratified by BMT status (**Figure 3**). In the AOF example, the number of predictors is small (all candidate AOF risk predictors were listed in step 2.1) in comparison to the size of our study sample/cases. Therefore,

it is possible to consider all possible combinations of predictors and interactions without concern of overfitting.

## 2.3 Tuning hyperparameter

Search the space of hyperparameters and select the combination that gives the best performance on the validation set.

**Note:** The purpose of tuning hyperparameters is to adjust the settings of the algorithm to optimize the model performance with respect to a specific data set. Usually, the values of hyperparameters are set with experience, or by doing a grid search with a wide range of values.

**Note:** Logistic regression does not have hyperparameters.

2.3.1 Support vector machines：test all three hyperparameters. They are linear, polynomial and Gaussian kernels.

2.3.2 Random Forest: test the number of trees in the forest, the number of predictors considered for splitting at each leaf node, and the minimal number of data points in a terminal node (i.e., node size).

## 2.4 K-fold cross-validation

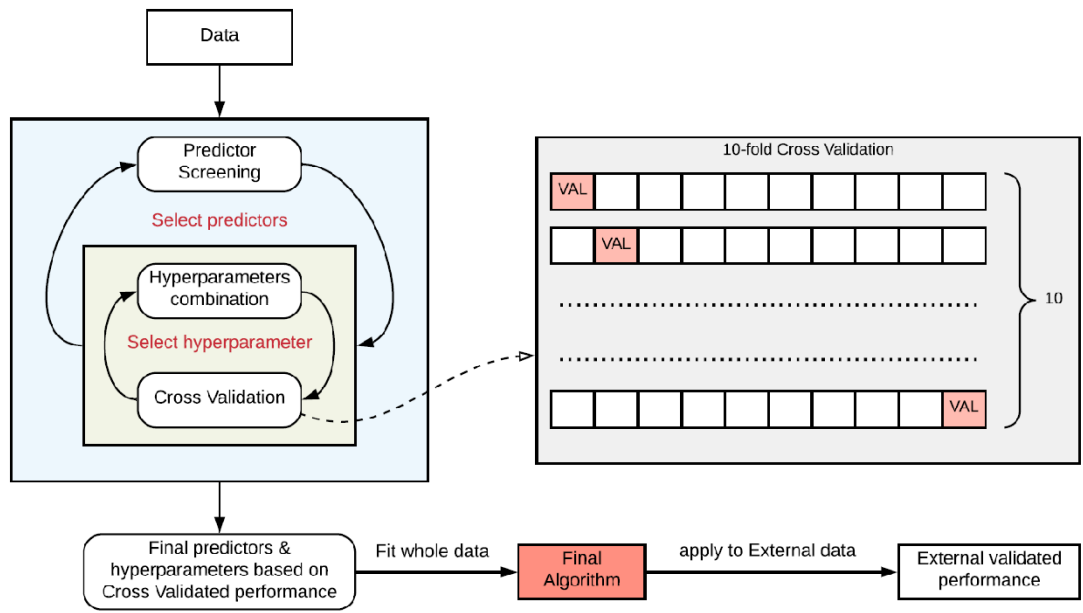

**Figure 2  workflow of model building**

**Note:** K-fold cross-validation was used for two purposes: 1) the tuning hyperparameters selection;

2) internal validation. The left panel of Figure 2 shows the process of 10-fold cross-validation.

2.4.1 Determine the value of K: In our example, K is set to be 10.

2.4.2 Randomly divide the total n labeled observations in the dataset into k=10 roughly equal sized parts .

2.4.3 Leave out one part of the data as the validation set and combine the remaining 9 parts as the training set, e.g. the first row in the right panel of **Figure 2**.

**Note:** in our example, we develop the algorithm such as logistic regression, random forest and support vector machine, using the training set, and "predict" AOF risk in the validation set, pretending the data is unlabeled.

2.4.4 Repeat 2.3.3 as illustrated in **Figure 2**.  Each time, one part of the data serves as the validation set and the risk of AOF on the validation set is predicted.  After repeating the process 10 times, combine all predicted risks and compare them to the observed outcomes to assess model performance as described in step 2.5.

**2.5 Performance assessment**

Validate the developed algorithms internally and then select the best modelling process to fit the whole data set to get the final model. Apply the final model to the external data set and report performance. Plot ROC, PR and calibration curves to visually assess the model performance. Estimate areas under the ROC and PR curves.

2.5.1 Internal validation: Plug in the observed AOF status data and the predicted risk of AOF to the "APBinary" function from the "APtools" package in R, which estimates the AUC values, AP values, and their 95% confidence intervals.  Create R functions to plot ROC, PR and calibration curves.

**Note:** There are different levels and methods for examining calibration.[12] To plot the calibration curve, we ranked subjects by their predicted risk from largest to smallest. The y-axis corresponds to the cumulative sum of the AOF risk (continuous) and the x-axis corresponds to the cumulative number of AOF survivors (discrete). We then scale both x and y-axis, dividing the values by the total number of AOF cases in the sample. A well-calibrated curve should be close to the 45° line.

2.5.2 External validation: Send the selected best model to the SJLIFE study team.

**Note:** For external validation, our collaborators in SJLIFE study team independently calculated the predicted risks for eligible subjects in the SJLIFE cohort. They estimated the AUC and AP values and calibration curve as described above using the predicted risks and observed AOF status.

**REPRESENTATIVE RESULTS:**

Random Forest automatically models complex interaction structures in data.[11] We plot the estimated risk of AOF from the random forest algorithm by inspecting its relationship with age at diagnosis stratified by BMT status, while holding other predictor values constant (**Figure 3**). It shows that the risk increases with age at diagnosis, regardless of BMT status. But the risk increases faster in survivors who were treated with BMT than those who were not, indicating that age modifies the effect of BMT status on AOF risk, i.e. there is an interaction between age at diagnosis and BMT status.

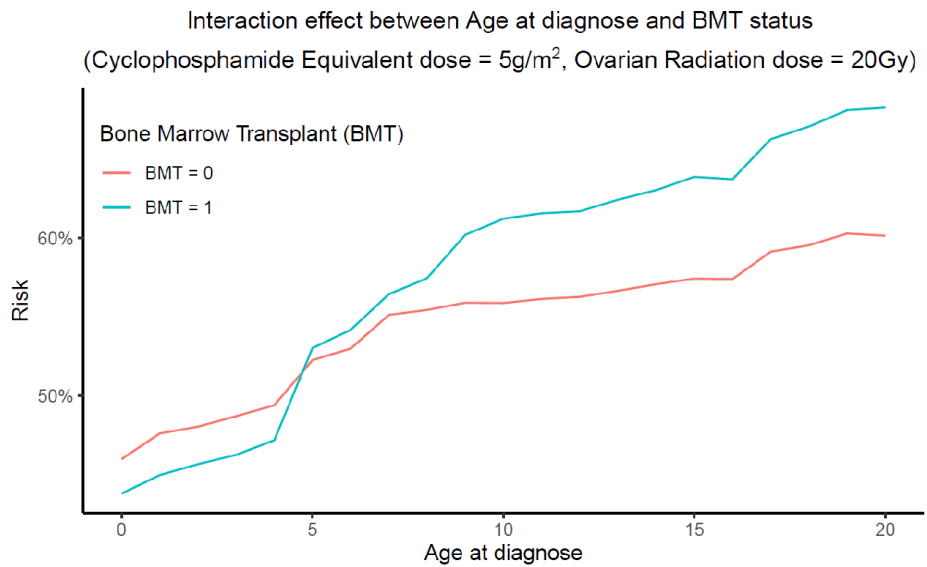

**Figure 3 Interaction effect between age at diagnose and Bone Marrow Transplant (BMT) status.**

Following the cross validation, the best predictors' combination and hyperparameters for each algorithm were identified. Assessment of the algorithm performance using a 10-fold cross validation is shown in **Table 1** and **Figures 4-6.** The estimates in the Table and curves in the Figures differ slightly from those published in our AOF risk prediction paper[4] because we used different internal validation approaches. As the three algorithms have very similar performance, the logistic regression model was selected for external validation because of the transparency and interpretability of the logistic model compared with the other two algorithms. The predictors included in the logistic model were minimum ovarian radiation dose, BMT status, age at cancer diagnosis, and CED value.

To examine the external validity, the chosen logistic regression algorithm was sent out to be externally validated using the SJLIFE cohort data. The results of external validation are shown in **Figures 7-9**, compared with the internal validation results. Our prediction algorithm achieved higher AUC and AP values in the external data than the internal curves, indicating that the prediction performance is robust. In the calibration curves, the cumulative observed cases are plotted against the cumulative predicted cases. While the internal validation calibration curve closely follows the diagonal line and is deemed calibrated, there is evidence that it is not as well calibrated in the external cohort because a deviation between the observed and predicted risks is observed.

**Table 1 Estimated AUC and AP values for the three algorithms from 10-fold cross-validation.**

|  | Logistic Regression | Random Forest | Support Vector Machines |
|---|---|---|---|
| AUC (95% Confidence interval) | 0.82 (0.78, 0.85) | 0.81 (0.78, 0.84) | 0.80 (0.77, 0.83) |
| AP (95% Confidence interval) | 0.50 (0.44, 0.56) | 0.47 (0.41, 0.53) | 0.48 (0.42, 0.53) |

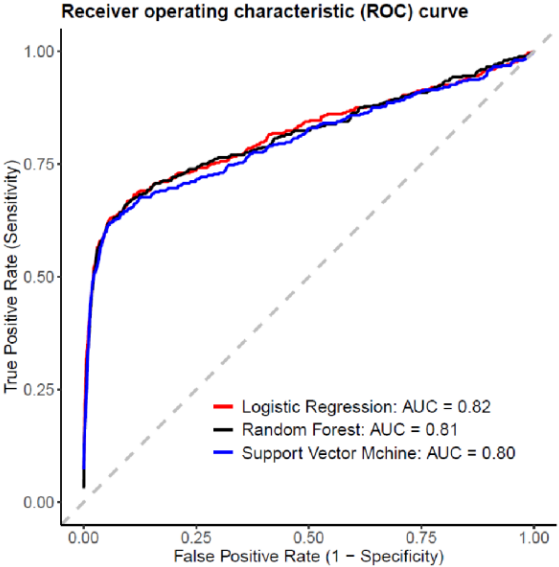

**Figure 4 Receiver operating characteristic (ROC) curves for three algorithms.**

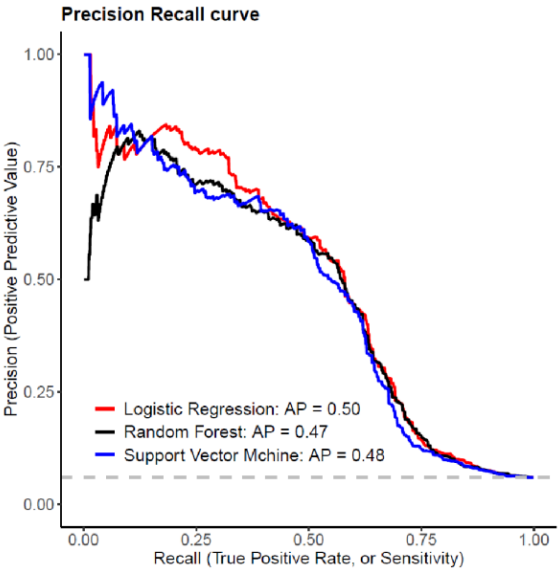

**Figure 5 Precision-recall curves for three algorithms.**

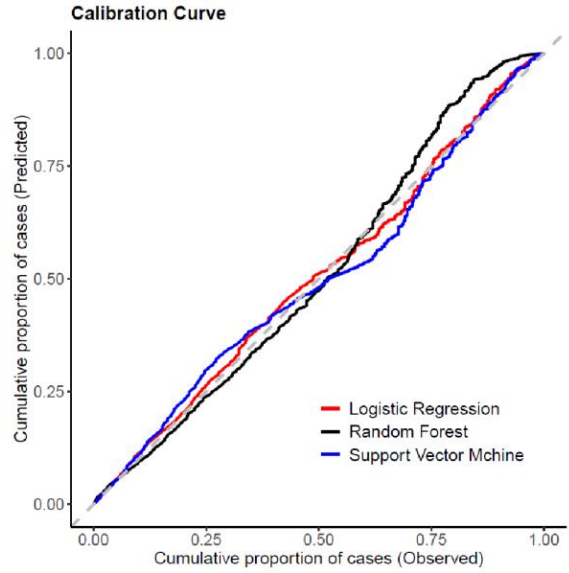

**Figure 6 Calibration curves for three algorithms.**

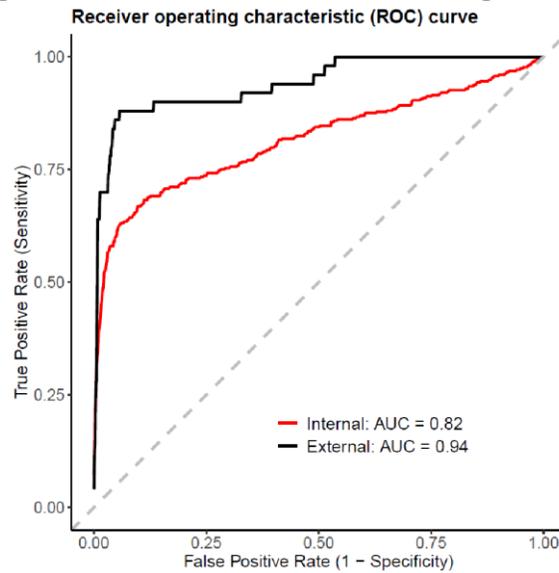

**Figure 7 ROC curves for logistic regression on internal and external data.**

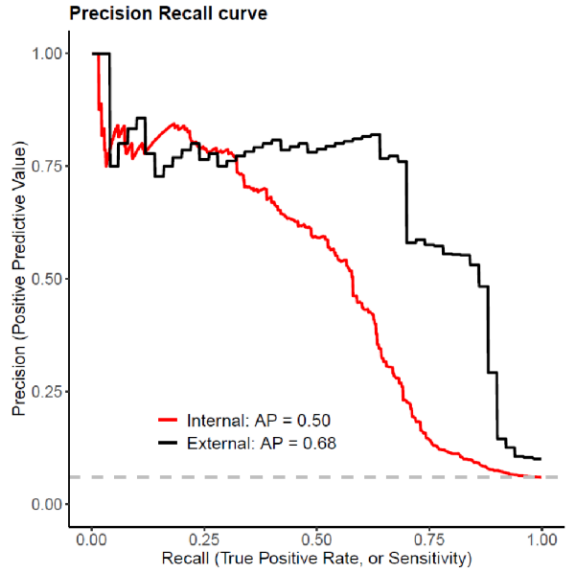

**Figure 8 Precision-Recall curves for logistic regression on internal and external data.**

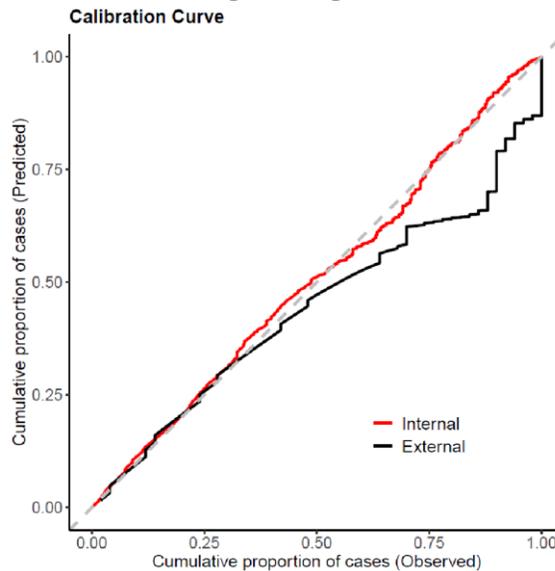

**Figure 9 Calibration curves for logistic regression on internal and external data.**

**FIGURE AND TABLE LEGENDS:**

**Figure 1 Entity Relationship Diagrams.** Each box represents a raw data set. PK: "Primary Key" is a field in a table which uniquely identifies each row/record in a data set. FK: "Foreign Key" is a key used to link two data sets together

**Figure 2 Framework of algorithm development.** The general workflow of algorithm development in predictive modelling. VAL stands for validation.

**Figure 3 Interaction effect between age at diagnose and Bone Marrow Transplant (BMT) status.**

**Figure 4** Receiver operating characteristic (ROC) curves for three algorithms.

**Figure 5** Precision-recall curves for three algorithms.

**Figure 6** Calibration curves for three algorithms.

**Figure 7** ROC curves for logistic regression on internal and external data.

**Figure 8** Precision-Recall curves for logistic regression on internal and external data.

**Figure 9** Calibration curves for logistic regression on internal and external data.

**Table 1** Estimated AUC and AP values for the three algorithms.

**DISCUSSION:**

High data quality is essential for a predictive model. Researchers should always carefully verify and ensure the quality of data prior to starting to build predictive algorithms for an outcome of interest. Otherwise, the algorithm is likely to be built on a shaky foundation and will lack internal and external validity. It is important to establish a data collection protocol that specifies the predictors, the outcome, and their measurements and implement it when collecting data. The data should be stored properly and securely with a data management protocol in place. Extensive exploratory data analysis provides a good initial check to identify the problematic data point and ensure data quality for the modeling step.

Interaction between predictors is an important consideration in predictive modeling. We risk model misspecification if an important interaction is not properly accounted for, which may lead to inaccurate predicted risk. However, it can be difficult to identify which predictors interact especially when there is a large collection of predictors. Several approaches can be used: 1) plausible interaction can be identified in discussion with content experts; 2) possible interaction may be identified by visually inspecting the outcome stratified by combinations of predictors; 3) taking advantage of computational algorithms such as a tree-based model, e.g. random forest, to identify the interactions between predictors (**Figure 3**).

It is critical to quantify the quality of the prediction of a predictive model. Many performance metrics can be used, e.g., mean square error and adjusted-$R^2$ for continuous outcomes, Brier score, AUC, AP and their extensions for binary and time-to-event outcome. Choosing a proper measurement is critical for the performance evaluation. In our study example, AOF has a low prevalence. The AP is a more suitable metric in this case because it gives more weight to correctly identify the high-risk patients compared to the AUC. Instead of building a model on one training set and evaluating on one test set for internal validation, better alternatives are data re-use strategies such as cross-validation, multiple train-test splitting, and bootstrap methods.[1] These strategies ensure the internal validity and reduce the variability of the estimates of the performance.

Tuning hyperparameters is an important aspect in the development of modern predictive algorithms. The most commonly used strategy to tune hyperparameters is to experiment with different combinations of hyperparameters using a grid search approach. We select the combination of hyperparameters that generates the best predictive performance. Usually, the software package has default values of hyperparameters, which can be used as a reference and starting point. The hyperparameter settings used in similar publications in the literature also provide good starting points.

In our study example, each of the three methods comes with its limitations. They make different assumptions about the data and simplify the complexity of the data to certain extents. Thus, some aspects of the underlying true data generation process could be lost. Subjected to a linear assumption and logit transformation, the logistic regression model is limited in its flexibility when modeling the relationship between continuous variables and binary outcomes. For modern algorithms, hyperparameter tuning can be time-consuming. To date, there is no good approach to improve efficiency of this step. Researchers often spend a lot of time on it. In addition, the results generated by modern algorithms such as support vector machines and random forest are "black-box", and not easily interpretable.

**ACKNOWLEDGMENTS:**
The publication of this manuscript has been supported by the Canadian Institutes of Health Research (FRN: 148693 to YY).

**DISCLOSURES:**
The authors have nothing to disclose.